\documentclass[manuscript]{emulateapj}

\newcommand{\xmm}{{\sl XMM-Newton}}

\begin{document}

\title{The {\it XMM-Newton} long look of NGC\,1365: 
lack of a high/soft state in its ultraluminous X-ray sources}

\author{
R.~Soria\altaffilmark{1}, 
G.~Risaliti\altaffilmark{2,3},
M.~Elvis\altaffilmark{2},
G.~Fabbiano\altaffilmark{2},
S.~Bianchi\altaffilmark{4},
Z.~Kuncic\altaffilmark{5}
}

\altaffiltext{1}{Mullard Space Science Laboratory, 
University College London, Holmbury St Mary, 
Dorking, Surrey RH5 6NT, United Kingdom; {\tt roberto.soria@mssl.ucl.ac.uk}}
\altaffiltext{2}{Harvard-Smithsonian Center for Astrophysics, 
60 Garden st., Cambridge, MA 02138, USA} 
\altaffiltext{3}{INAF--Osservatorio di Arcetri, Largo E.~Fermi 5, 
Firenze, Italy}
\altaffiltext{4}{Dipartimento di Fisica, Universita' degli Studi 
``Roma Tre'', via della Vasca Navale 84, I-00146 Roma, Italy}
\altaffiltext{5}{Sydney Institute for Astronomy, School of Physics, 
The University of Sydney, NSW 2006, Australia}

\begin{abstract}

Based on our long ($\sim 300$ ks) 2007 {\it XMM-Newton} 
observation of the Seyfert galaxy NGC\,1365, 
we report here on the spectral and timing behaviour 
of two ultraluminous X-ray sources, 
which had previously reached isotropic X-ray luminosities 
$L_{\rm X} \approx 4 \times 10^{40}$ erg s$^{-1}$ 
($0.3$--$10$ keV band). 
In 2007, they were in a lower state 
($L_{\rm X} \approx 5 \times 10^{39}$ erg s$^{-1}$, 
and $L_{\rm X} \approx 1.5 \times 10^{39}$ erg s$^{-1}$ 
for X1 and X2, respectively). Their X-ray spectra 
were dominated by power-laws with photon indices 
$\Gamma \approx 1.8$ and $\Gamma \approx 1.2$, respectively. 
Thus, their spectra were similar to those at their outburst peaks. 
Both sources have been seen to vary by a factor 
of 20 in luminosity over the years, but their spectra 
are always dominated by a hard power-law;
unlike most stellar-mass BHs, they have never 
been found in a canonical high/soft state dominated 
by a standard disk. The lack of a canonical high/soft state 
seems to be a common feature of ULXs. We speculate that the different 
kind of donor star and/or a persistently super-Eddington 
accretion rate during their outbursts may prevent 
accretion flows in ULXs from settling 
into steady standard disks.

\end{abstract}


\section{Introduction}

One of the hallmarks of accretion onto stellar-mass 
black holes (BHs) 
is a pattern of state transitions, initially discovered 
from X-ray spectral studies (starting 
from Cygnus X-1: \citet{tan72}) 
and later understood as fundamental changes 
in the structure of the accretion flow, 
and recognized also in the radio and optical behavior.
The ``canonical'' scheme of BH accretion \citep{rem06,fen04,esi98} 
consists of at least three main states: low/hard, high/soft 
and very high. The low/hard state is dominated 
by a power-law spectrum in the X-ray band, probably produced 
by inverse-Compton scattering of soft seed photons \citep{sun80}. 
The location and geometry of the scattering region is still 
unclear: it has been suggested that it could be 
a hot corona above the cold disk \citep{haa93,pou96}, 
an advection-dominated accretion flow (ADAF) \citep{nar94,esi97}, 
a fast jet or the base-of-the-jet \citep{fen04}, 
or a hot, centrifugally-supported boundary layer 
or post-shock region due to sub-Keplerian inflows \citep{cha95,cha97}. 
The high/soft state is dominated 
by thermal X-ray emission; most of the accretion power 
is radiated by an optically-thick, geometrically-thin disk 
\citep{ss73}. 
The very high state, observed when a source approaches its 
Eddington luminosity, is still poorly understood; 
it is again dominated by a power-law component 
in the X-ray band (usually steeper than in the low/hard state, 
\citet{rem06}), together with a thermal disk component 
that becomes relatively less important as the luminosity 
increases. The power-law component is thought 
to be produced by the Comptonization of soft X-ray photons 
from the inner disk region \citep{don06}. 
The spectrum of the disk component itself may be heavily modified 
by electron scattering, energy advection and radiation trapping, 
in the very high state (slim disk model: \citet{wat01}). 
In the simplest approximation, an accreting BH 
is in the radiatively-inefficient low/hard state 
when the mass accretion rate is less than 
a few per cent of the Eddington accretion rate 
(i.e., the accretion parameter 
$\dot{m} \equiv \dot{M}/\dot{M}_{\rm Edd} 
\approx \left(0.1 c^2\dot{M}\right)/L_{\rm Edd} \la 0.01$).
At higher accretion rates ($0.01 \la \dot{m} \la 0.5$), 
accreting BHs are expected to switch to 
the radiatively-efficient, disk-dominated 
high/soft state and then to the very high state, 
until the standard disk becomes unstable (at $\dot{m} \ga 1$). 
Another fundamental property, revealed by radio observations,  
is the presence in the low/hard state of a steady jet 
which may carry most of the outgoing accretion 
power; the jet is suppressed in the high/soft state 
\citep{mer03,fen04}; strong X-ray and radio flares 
are instead observed in the very high state \citep{rem06}.

One of the interesting properties of the standard disk 
spectrum in the high/soft state is that we expect 
a simple relation between the disk luminosity,  
the disk peak temperature, and the BH mass 
\citep{mak86,mak00}: 
$L_{\rm disk} \approx 4\pi R_{\rm in}^2 \sigma T_{\rm in}^4 
\sim M_{\rm BH}^2 T_{\rm in}^4$. In this formula,  
the inner-disk radius $R_{\rm in} \approx 6 R_{\rm g}$ 
for a non-rotating BH, where the gravitational radius 
$R_{\rm g} = M_{\rm BH}$ in geometric units. This relation has been 
observationally confirmed for many Galactic BHs, 
and the BH masses inferred with this method 
are generally in agreement (within a factor of 2) 
with direct kinematic measurements.
Another useful property of the high/soft state 
is the simple parameterization of disk luminosity 
and temperature with the accretion rate: 
for a fixed BH mass, $L_{\rm disk} \sim \dot{m}$, 
$T_{\rm in} \sim \dot{m}^{1/4}$ \citep{ss73}. 

The canonical state classification has also been applied 
to AGN, justified by the self-similar nature of many physical 
processes near a BH horizon, provided that  
characteristic sizes, timescales and disk temperatures 
are properly rescaled by the BH mass \citep{mac03,mer03,jes05,mar05}.
In this scheme, low-luminosity AGN and LINERs may represent 
the low/hard state, typical AGN and Seyferts 
may be in the high/soft state, and the most powerful 
quasars and FR-II radio galaxies 
may be in the very high state.

\begin{table*}
\begin{center}
\caption{Log of the 2007 {\it XMM-Newton} observations of NGC\,1365}
\begin{tabular}{lcccccr}
\hline \hline\\
Revolution & Obs ID & Instrument & Start time & End time  
     & Live time (ks) & Clean GTI (ks) \\[3pt]
\hline\\
1384  &  0505140201 & MOS & 2007-06-30 at 07:08:45  
  & 2007-07-01 at 18:51:55  
  & 126.7  & 96.6  \\
&& pn & 2007-06-30 at 07:31:03  
  & 2007-07-01 at 18:48:52   & 110.8  & 78.4 \\[3pt]
1385  &  0505140401 & MOS & 2007-07-02 at 07:07:24  
  & 2007-07-03 at 18:23:52  
  & 124.7  & 123.3  \\
&& pn & 2007-07-02 at 07:39:43  
  & 2007-07-03 at 18:35:02   & 109.2  & 107.9 \\[3pt]
1386  &  0505140501 & MOS & 2007-07-04 at 08:08:58  
  & 2007-07-05 at 18:27:14  
  & 121.7  & 92.8  \\
&& pn & 2007-07-04 at 08:36:56  
  & 2007-07-05 at 18:24:16   & 105.8  & 55.9 \\[3pt]
\hline
\end{tabular}
\end{center}
\end{table*}

Ultraluminous X-ray sources (ULXs) are accreting BHs 
with observed fluxes corresponding to isotropic X-ray 
luminosities up to $\sim$ a few $10^{40}$ 
erg s$^{-1}$, well above the Eddington limit 
for the kind of stellar BHs known in our Local Group 
(\citet{rob07} for a recent review). 
It is still widely debated whether this is 
due to extremely heavy stellar BHs, with masses 
$\sim 30$--$70 M_{\odot}$ \citep{pak02}, 
or intermediate-mass BHs \citep{mil04}, 
or collimation of the outgoing radiation \citep{kin01,kin08}, 
or super-Eddington luminosity \citep{beg02,beg06,ohs07};
or perhaps a combination of some of those factors. 
In other words, it is still unclear whether ULXs 
represent a different physical class of accreting BHs 
(perhaps with their own sequence of canonical states), 
or a different accretion state of ``ordinary'' stellar-mass BHs.

In the absence of direct kinematic measurements of BH masses 
in ULXs (mainly because of the faintness of their optical 
counterparts, at typical distances $\ga$ a few Mpc), 
X-ray spectroscopy could in principle provide strong constraints, 
by analogy with the properties of canonical spectral 
states in stellar-mass BHs. 
For example, if we observe a ULX in different spectral 
states over time, we could estimate its luminosity 
at the transition between low/hard and high/soft state, 
or the maximum luminosity in the low/hard state, 
and the fitted disk luminosity and temperature 
in the high/soft state. Such arguments were used 
for example by \citet{win06} and \citet{mil04}, 
in support of BHs with masses 
$\ga 10^3 M_{\odot}$. However, this method 
relies on the assumption that ULXs follow the same 
pattern of canonical spectral states; specifically, 
that they will switch from the low/hard to the 
high/soft state at a luminosity $\sim$ a few per cent 
of $L_{\rm Edd}$, and that they are dominated by 
a standard disk in the high/soft state. 

Until now, it has been difficult to test these assumptions.
The main reason why we still know very little about 
spectral states transitions in ULXs is that 
none of these sources has had the kind of daily 
or weekly X-ray monitoring that is possible 
for Galactic BHs. Even the brightest ULXs 
are usually observed only sporadically, every 
few months or years. This makes it difficult 
for us to understand the sequence and timescale 
of their spectral state transitions. 
Nonetheless, for a few ULXs there are now 
enough datapoints over the years, so that 
we can try and compare their behaviour with those 
of transient Galactic BHs.

NGC\,1365 X1 is one of such rare sources, detected by  
{\it ROSAT}, {\it ASCA}, {\it Chandra}, {\it Swift} and {\it XMM-Newton} 
at various levels of activity between 1993 and 2007 
\citep{kom98,sor07,str08}, with peak luminosities 
$\approx 3$--$4 \times 10^{40}$ erg s$^{-1}$ 
in 1995 and 2006 \citep{sor07}.
The nearby transient ULX X2 also reached a luminosity 
$\approx 4 \times 10^{40}$ erg s$^{-1}$ in 2006 
\citep{str08}, but was much fainter 
or undetected during the earlier observations.
Their host galaxy is the most luminous spiral in the Fornax cluster, 
[type SBb(s)I: \citet{san81}] with a total gas mass 
$\approx 3 \times 10^{10} M_{\odot}$, a kinematic mass  
$\approx 3.6 \times 10^{11} M_{\odot}$, and a star formation rate  
$\approx 10 M_{\odot}$ yr$^{-1}$ \citep{lin99,rou01}. 
The Cepheid distance to NGC 1365 is $19 \pm 1$ Mpc \citep{fer00}; 
at that distance,  1 arcsec $= 92$ pc.
In this paper, we report the results of our $\approx 300$-ks 
{\it XMM-Newton} observation, 
split over three orbits between 2007 June 30 and 2007 July 05 (Table 1 
and Section 2 for details). 
We interpret the X-ray properties 
of the two ULXs (in particular, their hard power-law spectra) 
in the context of their long-term behaviour.

\begin{figure}[t]
\includegraphics[angle=0,width=1\columnwidth]{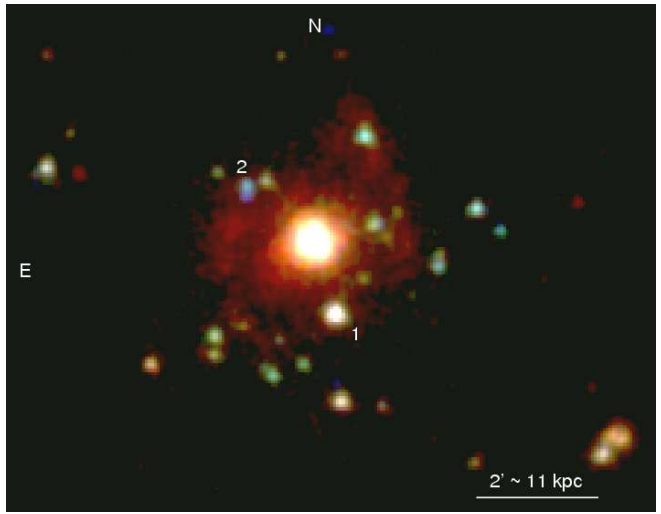}\\
\figcaption{{\it EPIC}-MOS true-color image of the central 
region of NGC\,1365. Red $= 0.3$--$1$ keV; green $=1$--$2$ keV; 
blue $=2$--$10$ keV. The two ULXs discussed in this paper 
are marked as ``1'' and ``2''.}
\label{fig:figure1}
\vspace{0.3cm}
\end{figure}

\section{Observations and data analysis}

{\it XMM-Newton} observed NGC\,1365 on 2007 June 30--July 05 
(Table 1); the Principal Investigator was Guido Risaliti. 
The prime instrument was the European Photon Imaging Camera (EPIC), 
in the Full Window mode, Medium filter. 
See Figure 1 for a true-colour image of the combined MOS dataset.
The primary target was the highly variable Seyfert nucleus, 
on which we report elsewhere (Risaliti et al.~2008, in prep.).
We used the \xmm\ Science Analysis System ({\footnotesize{SAS}}) 
version 7.1.0 to process and filter the event files and 
extract lightcurves and spectra.
Several background flares occurred during the three 
observations. We tried various rejection thresholds 
to increase the signal-to-noise for the non-nuclear sources; 
in particular, to improve the signal-to-noise of X1 and X2 
at photon energies $\ga 5$ keV (the two sources have 
similar fluxes in that energy range). Instead, the nucleus 
is brighter, and therefore much less affected by the 
flares, except for a particularly strong, short one at the end 
of the second observation. In the end, for our spectral analysis
we chose to retain good-time-intervals of 
78.4 ks,  107.9 ks, and 55.9 ks for the EPIC-pn 
in the three observations, and 
96.6 ks, 123.3 ks, and  92.8 ks for the EPIC-MOS (Table 1).
For the timing analysis, we used the whole live exposure time, 
although of course the intervals with high or flaring 
background give lower signal-to-noise bins in the lightcurve.

The extraction of spectra and lightcurves for X1 
was made difficult by the unfortunate location 
of the source on a chip gap in the pn, leading 
to a loss of about half the counts in every 
exposure. The ancillary response function
is supposed to take into account the reduced 
effective area of the source region; however, 
because the point spread functions of hard and 
soft photons are slightly different, a chip gap 
may in principle introduce a change in the spectral 
shape. Fortunately, X1 is not on a chip gap in the MOS 
images. We fitted separately and compared 
the MOS and pn spectra, and verified that they are 
consistent with each other. Thus, we conclude  
that the chip gap may have decreased the signal-to-noise 
achievable in principle, but has not significantly 
affected the spectral results.

The radius of the source extraction regions 
was $20\arcsec$ for X1 (the brightest non-nuclear source), 
and only $8\arcsec$ for X2, to reduce contamination 
from a nearby source of similar brightness (X2 was much 
fainter than in 2006). Background extraction 
regions were chosen around the source regions, in a suitable 
way to avoid contamination. After building 
response and ancillary response files with 
{\footnotesize{rmfgen}} and {\footnotesize{arfgen}}
we used {\footnotesize XSPEC} \citep{arn96} for spectral analysis.
To improve the signal-to-noise ratio, we co-added 
the EPIC pn and MOS spectra from each observation,
and for all the observations, 
with suitably averaged response functions,
using the method of \citet{pag03}.

\section{Main Results}

\subsection{Time variability of X1}

We extracted background-subtracted MOS and pn lightcurves 
of X1 in the three observations, and analyzed them 
with standard {\footnotesize FTOOLS} tasks, 
with various choices of binning.
We did not find any significant variability between 
the three observations, nor within a single observation.
For example, for the 2007 July 02--03 EPIC-pn observation 
(Rev.~1385), which has the highest signal-to-noise, 
we find a $\chi^2$ probability of constancy $> 0.9$ 
(Figure 2). We do not find any significant features 
in the power density spectra, either. 

Over the past fourteen years, X1 has shown strong 
variability, with two major outbursts detected 
by {\it ASCA} in 1995 and {\it Chandra} in 2006 
(Figure 3). In addition to the luminosity datapoints 
already summarized in \citet{sor07}, we have also 
analyzed a series of short {\it Swift} monitoring 
observations, carried out on 2006 July 21--24 
(three months after the {\it Chandra} observation), for 
a total exposure time of 19.4 ks. We detect 23 net counts 
with the X-Ray Telescope (count rate $\approx 0.0012$ count 
s$^{-1}$), corresponding to an emitted luminosity 
$\approx (2.3 \pm 0.5) \times 10^{39}$ erg s$^{-1}$ 
in the $0.3$--$10$ keV band (Figure 3), assuming a power-law 
model with photon index $1.8$ and 
$N_{\rm H} = 5 \times 10^{20}$ cm$^{-2}$. 


\begin{figure}
\includegraphics[angle=270,width=1\columnwidth]{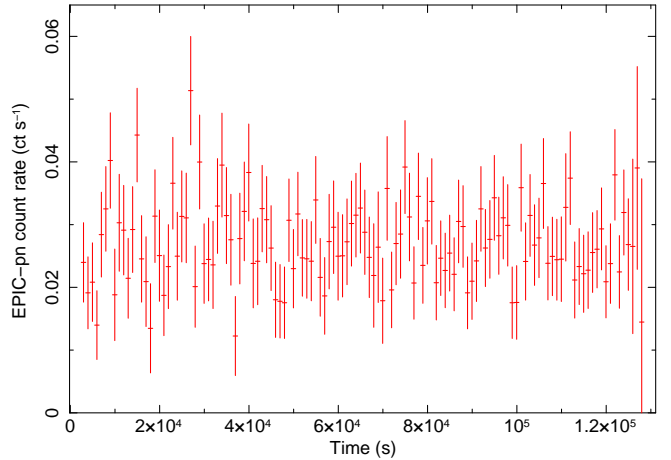}
\figcaption{EPIC-pn lightcurve of X1 (binned to 1000 s) 
from the second {\it XMM-Newton} observation of 2007 
(Rev.~1385).}
\label{fig:figure2}
\vspace{0.3cm}
\end{figure}

\begin{figure}
\includegraphics[angle=270,width=1\columnwidth]{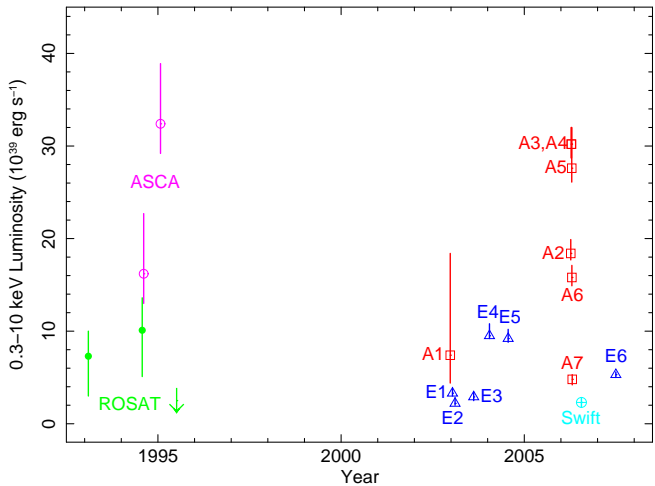}
\figcaption{Long-term X-ray lightcurve of X1: 
unabsorbed luminosities in the $0.3$--$10$ keV band. 
Green filled circles are the luminosities extrapolated 
from {\it ROSAT}; open magenta circles from {\it ASCA}; 
red squares from {\it Chandra}; blue triangles from {\it XMM-Newton}; 
an open cyan circle from {\it Swift}. 
For details on the observations from 1993 to early 2006, 
see \citet{sor07}.
The peak {\it ASCA} luminosity 
$\approx 3$--$4 \times 10^{40}$ erg s$^{-1}$ is based 
on our own re-analysis of the data \citep{sor07}, 
and is about half of the value often quoted 
in the literature.}
\label{fig:figure3}
\vspace{0.3cm}
\end{figure}

\subsection{Spectral analysis: X1 is a power-law ULX}

Firstly, we fitted and compared the MOS spectra of X1 from 
each {\it XMM-Newton} observation during the three 
consecutive revolutions of 2007 June 30 -- July 05. 
We find that the three spectra are consistent with each other 
with regard to observed flux, absorbing column density 
and spectral shape (Figure 4). All three spectra 
are well fitted by an absorbed power-law model 
with total absorbing column density 
$\sim 5 \times 10^{20}$ cm$^{-2}$ 
and power-law photon indices $\approx 1.7$--$1.8$, 
consistent within the 90\% confidence limit 
of each dataset. Hence, we are justified in adding 
the EPIC spectral data from all three observations 
to increase the signal-to-noise ratio.

The combined {\it XMM-Newton}/EPIC spectrum is well fitted 
($\chi^2_{\nu} = 1.03(156.2/152)$) by a simple 
power-law model (see the top panel of Figure 5, 
and Table 2 for the best-fitting parameters). 
From this fit, we estimate an unabsorbed, 
isotropic luminosity $L_{\rm X} \approx 5.3 \times 10^{39}$ 
erg s$^{-1}$ in the $0.3$--$10$ keV band.
Some ULXs often show a curvature in their X-ray spectra 
(see Section 4), to the point that it is sometimes difficult 
to distinguish between a power-law, a cut-off power-law 
and a disk-blackbody model. It is clearly not the case here: 
we show the best-fitting disk-blackbody model in the bottom 
panel of Figure 5 ($\chi^2_{\nu} > 2$), as a graphic evidence 
of the lack of spectral curvature. For the same reason, 
even a bremsstrahlung model, although slightly less curved 
than a disk-blackbody, does not fit the data well; the best-fitting 
model has $\chi^2_{\nu} = 1.13(172.3/152)$, for $kT = (5.6 \pm 0.6)$ keV.

A power-law is the best-fitting 1-component phenomenological
model, and it makes no assumptions on the underlying 
physical process. It is of course possible to fit the same 
spectrum equally well with other, more complex physical models, provided 
that they look exactly like a power-law in the observed band. 
For example, we tried the Comptonization model {\it comptt} 
and found that we could get equally good fits 
by fixing the seed-photon temperature at any value 
$kT_0 \la 0.1$ keV, if the coronal temperature $kT_e \ga 2$ keV 
and the optical depth $\tau \ga 0.5$. The best-fitting {\it comptt} model 
has $\chi^2_{\nu} = 1.03(155.2/151)$ for $kT_e \approx 2.9$ keV 
and $\tau \approx 6.2$ (warm, optically-thick corona). 
But we also get a statistically equivalent fit 
($\chi^2_{\nu} = 1.04(157.8/152)$) for example for $kT_e \ga 50$ keV 
and $\tau \equiv 0.5$ (hot, optically-thin corona).
Both models are indistinguishable from a power-law, 
with the available data.

We then tried adding a {\it diskbb} component (soft excess) 
to the power-law: the best-fitting model has $\chi^2_{\nu} = 1.02(153.2/150)$
for $kT_{\rm in} = 0.35^{+0.17}_{-0.13}$ keV and 
photon index $\Gamma = 1.66^{+0.15}_{-0.11}$, 
but this is not a statistically-significant improvement 
with respect to the simple power-law model.
From the upper limit to the {\it diskbb} normalization, 
we find that, if there is an additional disk-blackbody component, 
it does not contribute more than $\approx 3.6 \times 10^{38}$ 
erg s$^{-1}$ in the $0.3$--$10$ keV band 
($\approx 5.4 \times 10^{38}$ erg s$^{-1}$ bolometric), 
that is $< 7$\% of the power-law contribution.

Finally, we looked for a possible break 
or cut-off in the power-law at high energies.
An exponential cut-off does not improve the fit, because 
it introduces excessive curvature in the model.
A break at the energy $E_{\rm b} = 7.5 \pm 0.9$ keV
gives $\chi^2_{\nu} = 1.00(150.3/150)$, which is, formally, 
a marginally significant improvement from a simple power-law 
fit, and confirms the visual impression of a spectral downturn 
(Figure 5, top panel). However, given the low signal-to-noise 
level at those channel energies, and the uncertainties in the background 
subtraction (contaminated by the strong AGN emission above $\approx 6$ keV), 
we cannot attribute any physical meaning to this possible break. 
Nonetheless, we recall that there was also 
marginal evidence of a break at $\approx 6$ keV in one 
of the {\it XMM-Newton} observations of 2004 \citep{sor07}.

The power-law photon index $\Gamma \approx 1.8$ (Table 2) is consistent 
with the indices fitted to the {\it Chandra} spectra during 
the 2006 April outburst (see Table 2 for the best-fitting 
parameters of the combined 2006 April 12--15 spectrum), 
and also with the indices fitted to the {\it ASCA}, 
{\it Chandra} and {\it XMM-Newton} spectra over 1995--2003. 
In the 2004 {\it XMM-Newton} spectra, instead, X1 
showed a soft excess (inner disk temperature 
$kT_{\rm in} \approx 0.3$--$0.4$ keV) contributing 
about 25--30 per cent of the X-ray luminosity, 
in addition to a dominant (and harder) power-law 
component with a high-energy break. 
At that time, the unabsorbed luminosity
was $\approx 10^{40}$ erg s$^{-1}$ \citep{sor07}.

The column density in the 2007 spectrum is consistent 
with the values previously fitted to the {\it XMM-Newton} 
spectra when X1 had X-ray luminosities $\la 10^{40}$ erg s$^{-1}$.
The neutral column density was instead a few times higher 
during the 2006 outburst, when a single-component 
power-law model was used \citep{sor07}. 
As a further check that the decrease in column density 
between 2006 and 2007 is real and not due to the appearance 
of an underlying soft component, we fitted the 2007 spectrum
with an additional {\it diskbb} component (leaving its temperature 
and normalization as free parameters) and $N_{\rm H}$ 
fixed at the 2006 value. We obtain $\chi^2_{\nu} = 1.15(173.9/151)$ 
which is significantly worse than the low-column best fit.
Besides, there was also strong ionized absorption at the peak 
of the 2006 outburst \citep{sor07}, which is absent 
from the 2007 spectra. 

We also re-examined the 2006 data to test the opposite  
scenario: the possibility that the column density in 2006 
was as low as in 2007 or at any other epoch. We started from 
the best-fitting model ($\chi^2_{\nu} = 0.72(105.5/146)$) 
to the 2006 outburst summarized in Table 7 of \citet{sor07}, but this time 
we imposed $N_{\rm H} \equiv 4.5 \times 10^{20}$ cm$^{-2}$. A single 
power-law model is no longer acceptable ($\chi^2_{\nu} = 1.06(155.5/147)$), 
because it does not reproduce the  low-energy downturn. 
A low-$N_{\rm H}$ disk-blackbody model is statistically worse 
than the best-fitting model ($\chi^2_{\nu} = 0.76(113.2/146)$): 
it can reproduce the low-energy downturn but, in that case, it 
overpredicts high-energy curvature; the best-fitting 
$kT_{\rm in} \approx 1.4$ keV
(a characteristic temperature that makes it difficult for 
{\it Chandra}/ACIS to distinguish between an absorbed power-law 
and a disk-blackbody model, because of its limited spectral coverage).
On the other hand, low-$N_{\rm H}$ Comptonization models provide 
statistically-equivalent fits to the high-$N_{\rm H}$ power-law model, 
if the seed photon temperature $\approx 0.3$ keV.
For example, using the Comptonization model {\it bmc}, 
we obtain $\chi^2_{\nu} = 0.72(104.5/145)$ with the best-fitting 
seed photon temperature $kT_0 = 0.27^{+0.03}_{-0.04}$ keV.
The same Comptonization models applied to the 2007 spectra require 
a lower seed photon temperature, $\la 0.1$ keV.
As for the ionized absorption component seen at the peak of the 2006 
outburst \citep{sor07}, we have not found it possible to eliminate it 
or replace it with any other simple spectral component.

\begin{table}
\begin{center}
\caption{Best-fitting parameters for the combined 
2007 {\it XMM-Newton}/EPIC spectrum 
of X1, compared with the 2006 peak outburst 
spectrum from {\it Chandra} data. 
Spectral model: {\tt phabs*phabs*power-law}. Errors are 90\%
confidence levels for 1 interesting parameter ($\Delta \chi^2 =2.7$). }
\vspace{0.2cm}
\begin{tabular}{lcr}
\tableline\tableline\\
Parameter & 2006 {\it Chandra} Value & 2007 {\it XMM-Newton} Value \\
\tableline\\
$N_{\rm {H,Gal}}$\tablenotemark{a} & $1.3 \times 10^{20}$ & $1.3 \times 10^{20}$\\[6pt]
$N_{\rm {H}}$ & $1.5^{+0.4}_{-0.3} \times 10^{21}$ & $4.3^{+1.5}_{-1.5} \times 10^{20}$\\[6pt]
$\Gamma$\tablenotemark{b}  & $1.74^{+0.12}_{-0.11}$ & $1.80^{+0.04}_{-0.05}$\\[6pt]
$N_{\rm {pl}}$\tablenotemark{c} &  $9.8^{+1.3}_{-1.1} \times 10^{-5}$ 
                  &  $1.9^{+0.1}_{-0.1} \times 10^{-5}$\\[6pt]
\tableline\\
$\chi^2$/dof & $0.93 (85.4/92)$ & $1.03 (156.2/152)$ \\[3pt]
\tableline\\
$f_{\rm 0.3-10}$\tablenotemark{d} &$5.2^{+0.5}_{-0.6} \times 10^{-13}$ 
       &$1.1^{+0.1}_{-0.1} \times 10^{-13}$\\[6pt]
$L_{\rm 0.3-10}$\tablenotemark{e} &$2.8^{+0.3}_{-0.3} \times 10^{40}$
        & $5.3^{+0.3}_{-0.3} \times 10^{39}$ \\[6pt]
\tableline\\
\end{tabular}
\vspace{-0.5cm}
\tablenotetext{a}{From \citet{K05}. Units of cm$^{-2}$.}
\tablenotetext{b}{Photon index.}
\tablenotetext{c}{Units of photons keV$^{-1}$ cm$^{-2}$ s$^{-1}$, at 1 keV.}
\tablenotetext{d}{Observed flux in the $0.3$--$10$ keV band; 
          units of erg cm$^{-2}$ s$^{-1}$.}
\tablenotetext{e}{Unabsorbed luminosity in the $0.3$--$10$ keV band; 
          units of erg s$^{-1}$.}\\
\end{center}
\vspace{0.2cm}
\end{table}

\begin{table}
\begin{center}
\caption{Best-fitting parameters for the combined 
2007 {\it XMM-Newton}/EPIC spectrum of X2, compared 
with the 2006 peak outburst spectrum from {\it Chandra} data. 
Spectral model: {\tt phabs*phabs*power-law}. Errors are 90\%
confidence levels for 1 interesting parameter ($\Delta \chi^2 = 2.7$). }
\vspace{0.2cm}
\begin{tabular}{lcr}
\tableline\tableline\\
Parameter & 2006 {\it Chandra} Value & 2007 {\it XMM-Newton} Value \\
\tableline\\
$N_{\rm {H,Gal}}$\tablenotemark{a} & $1.3 \times 10^{20}$ & $1.3 \times 10^{20}$\\[6pt]
$N_{\rm {H}}$ & $8.1^{+2.6}_{-1.9} \times 10^{21}$ & $1.2^{+0.7}_{-0.6} \times 10^{21}$\\[6pt]
$\Gamma$\tablenotemark{b}  & $1.23^{+0.25}_{-0.19}$ & $1.13^{+0.09}_{-0.10}$\\[6pt]
$N_{\rm {pl}}$\tablenotemark{c} &  $7.6^{+3.2}_{-2.1} \times 10^{-5}$ 
                  &  $2.6^{+0.6}_{-0.5} \times 10^{-6}$\\[6pt]
\tableline\\
$\chi^2$/dof & $0.70 (23.2/33)$ & $1.12 (110.8/99)$ \\[3pt]
\tableline\\
$f_{\rm 0.3-10}$\tablenotemark{d} &$6.8^{+1.2}_{-3.2} \times 10^{-13}$ 
       &$3.2^{+0.4}_{-0.6} \times 10^{-14}$\\[6pt]
$L_{\rm 0.3-10}$\tablenotemark{e} &$3.7^{+0.3}_{-0.3} \times 10^{40}$
        & $1.5^{+0.1}_{-0.1} \times 10^{39}$ \\[6pt]
\tableline\\
\end{tabular}
\vspace{-0.5cm}
\tablenotetext{a}{From \citet{K05}. Units of cm$^{-2}$.}
\tablenotetext{b}{Photon index.}
\tablenotetext{c}{Units of photons keV$^{-1}$ cm$^{-2}$ s$^{-1}$, at 1 keV.}
\tablenotetext{d}{Observed flux in the $0.3$--$10$ keV band; 
          units of erg cm$^{-2}$ s$^{-1}$.}
\tablenotetext{e}{Unabsorbed luminosity in the $0.3$--$10$ keV band; 
          units of erg s$^{-1}$.}\\
\end{center}
\vspace{0.2cm}
\end{table}

\subsection{X2: a transient, hard power-law ULX}

The transient ULX X2 (Figure 1) was serendipitously discovered 
during the 2006 {\it Chandra} observations 
\citep{str08}\footnote{It is not the same 
source designated as X2 in Fig.~2 of \citet{sor07}.}. 
It underwent an outburst with a rapid rise time 
(a factor of 7 in two days; \citet{str08}) 
at the same time as the unrelated ULX X1 was also in outburst. 
In our 2007 {\it XMM-Newton} observations, the source was much 
fainter, and much less variable; the observed flux is the same 
in the three observations, within $1\sigma$. 
Remarkably, the slope of the X-ray spectrum 
was similar in 2006 and 2007, well fitted 
by a hard power-law with $\Gamma \approx 1.1$--$1.2$ (Table 3 
and Figure 6). 
The column density in 2007 was $\approx 10^{21}$ cm$^{-2}$, 
an order of magnitude less than at the peak 
of the 2006 outburst.
The unabsorbed isotropic luminosity reached 
$\approx 3.7 \times 10^{40}$ erg s$^{-1}$ in 2006, 
and was down to $\approx 1.5 \times 10^{39}$ erg s$^{-1}$ in 2007.

As for X1, we tried fitting the 2007 {\it XMM-Newton} spectrum 
of X2 with alternative models, 
even though the lower signal-to-noise makes multi-component 
models even less constrained.
A {\it diskbb} model provides a formally acceptable fit 
($\chi^2_{\nu} = 1.15(113.6/99)$, compared with 
$\chi^2_{\nu} = 1.12(110.8/99)$ for the power-law model).
But the best-fitting temperature $kT_{\rm in} = 3.48^{+1.31}_{-0.66}$ keV
is unphysically high for a standard disk, and the normalization 
corresponds to an unphysically small inner radius 
$r_{\rm in} (\cos \theta)^{0.5} \approx 7$ km.
A bremsstrahlung model provides a formally acceptable fit 
($\chi^2_{\nu} = 1.14(112.6/99)$) for $kT \ga 48$ keV.
The Comptonization model {\it comptt} with seed photon temperature fixed 
at $kT_0 \equiv 0.1$ keV and optically thin corona ($\tau \equiv 0.5$)
gives an equivalent fit ($\chi^2_{\nu} = 1.12(111.3/99)$) for 
$kT_e \ga 250$ keV.
For an optically-thick corona ($\tau \equiv 10$), the best-fitting 
model has $kT_e = 2.73^{+0.66}_{-0.38}$ keV
($\chi^2_{\nu} = 1.11(109.7/99)$).
An exponential cutoff does not improve the power-law fit 
($\chi^2_{\nu} = 1.12(110.8/99)$), and the cutoff is constrained 
to be at energies $\ga 5.8$ keV. A break does not improve 
the power-law fit, either. Adding a soft disk-blackbody 
component to the power-law does not improve 
the fit: the best fitting model has a null normalization for 
{\it diskbb}, and an upper limit of $\approx 25$\% 
for the disk-blackbody contribution.
In short, as was the case for X1, also X2 can be fitted 
by various physical models, as long as they look exactly  
like a hard power law in the observed band.
We then fixed the absorbing column density to the 2006 
value (Table 3) and verified that it is not consistent 
with the 2007 data, even with the addition of 
an unconstrained disk-blackbody component to compensate 
for that. The best-fitting high-absorption model 
has $\chi^2_{\nu} = 1.25(122.5/98)$. 
Finally, we re-examined the 2006 data, 
and tried fitting them with the same low absorption as in 2007.
As for the case of X1, we find that Comptonization models 
with low absorption but high seed-photon blackbody temperature 
($kT_0 \approx 0.9$ keV) give fits that are only slightly worse 
than the high-absorption models. However, such scenario is too contrived, 
and unseen in any other accreting source. Hence, 
we conclude that the decrease in absorption at lower 
luminosities is physically significant for X2.

In previous years, X2 was not seen by {\it ASCA}. 
It was claimed \citep{str08} that it may 
have been marginally detected by {\it ROSAT}/HRI in 1995, 
at a count rate corresponding to a $0.3$--$10$ keV luminosity 
$\approx 4 \times 10^{39}$ erg s$^{-1}$, if we assume 
the spectral parameters of the 2007 {\it XMM-Newton} spectrum; 
however, this detection is doubtful, because of likely 
source confusion. X2 was also in a low state 
($L_{\rm X} \la 3 \times 10^{38}$ erg s$^{-1}$) during the 2002 
{\it Chandra} observation. It may be marginally detected 
(but confused with nearby, brighter sources) in 
the 2004 {\it XMM-Newton} observations, 
at a luminosity $\sim$ a few $10^{38}$ erg s$^{-1}$.
It is not detected ($L_{\rm X} \la 10^{39}$ erg s$^{-1}$) 
in the {\it Swift} observations of 2006 July.

\begin{figure}
\includegraphics[angle=270,width=1\columnwidth]{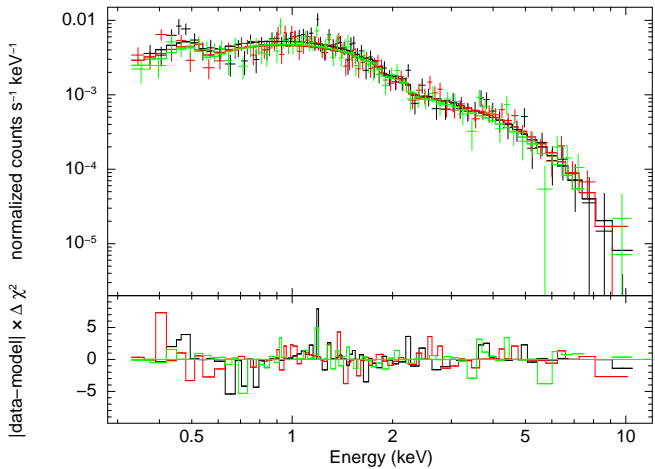}
\figcaption{Individual {\it EPIC}-MOS spectra of X1 
during the three observations in 2007, and respective 
$\chi^2$ residuals, for absorbed power-law fits. Black 
datatpoints and residuals are for Rev.~1384; red ones 
for Rev.~1385; green ones for Rev.~1386. The three spectra 
are statistically identical in flux normalization 
and spectral slope. 
}
\label{fig:figure4}
\vspace{0.3cm}
\end{figure}

\section{Discussion and Speculations}

\subsection{Power-law versus thermal disk spectra}

The strongly star-forming spiral galaxy NGC\,1365 contains 
many X-ray sources that have exceeded the ULX threshold 
at some stage in the past few years. Here we have focused 
on two sources that showed recent outbursts and great changes 
in flux; for one of them (X1) there is a long sequence 
of X-ray observations dating back to 1993.
The X-ray spectra of X1 and X2 appear always dominated 
by broad, power-law-like components with photon 
indices $\Gamma \approx 1.8 \pm 0.1$ and $\Gamma \approx 1.2 \pm 0.1$, 
respectively, both in their higher and lower states 
(even though their luminosities varied by a factor of 20). 
For X1, sometimes ({\it XMM-Newton} observations 
from 2004, \citet{sor07}) there is also a cool, 
thermal soft component 
in addition to the dominant, hard power-law component.
For both sources, a significant difference 
in the $0.3$--$10$ keV band bewteen 
spectra taken at higher and lower luminosity is the higher 
intrinsic column density in the higher state. 
We do not have enough elements to determine whether 
this is directly related to a higher mass inflow 
or outflow rate, or both.

Based on their phenomenological spectral appearance, 
one may be tempted to classify both ULXs
in the power-law-dominated low/hard state,
using the canonical definition of BH spectral states. 
Several other luminous ULXs (with 
$L_{\rm X} \la 3 \times 10^{39}$ erg s$^{-1}$) 
are well fitted by a single power-law of photon 
index $1 \la \Gamma \la 2$ \citep{ber08}: such  
slopes are consistent with and sometimes even harder 
than those measured in the low/hard state 
of Galactic BHs.
However, the first hint that the accretion state 
of X1 and X2 may not be the canonical low/hard state 
comes from their long-term variability:
both sources remained in the same hard state 
over the years, despite their luminosity changes.
This is different from the behaviour observed in most 
stellar-mass Galactic BHs, which recurrently switch 
between the low/hard and high/soft states. 
In particular, in most Galactic soft X-ray transients, 
outbursts tend to start 
in the hard state, but rapidly switch to the soft state 
(on a timescale of few days, during which the disk builds up); 
then, such sources spend the majority of their bright time 
(a few weeks or months) in the high/soft state, dominated 
by standard disk-blackbody emission with $L \sim T_{\rm in}^4$ 
and $0.1 \la L \la L_{\rm Edd}$.

This does not appear to be the case for ULXs: 
none of the most luminous or best studied sources 
has ever been unequivocally found in a canonical high/soft state. 
Some ULXs do have a convex spectrum 
\citep{mak00,fen05,sto06,tsu07,miz07,mak07,miy08}, 
with various degrees of curvature between a straight power-law and 
a disk-blackbody. But if they are formally fitted with 
a disk-blackbody model, their peak color temperature 
($kT_{\rm in} \sim 1.5$--$2.5$ keV) is much higher 
than expected for a standard disk, and implies 
a luminosity $L > L_{\rm Edd}$, inconsistent 
with the standard-disk model. One possible interpretation 
is that the emission is from a hot, super-Eddington 
slim disk \citep{wat01,mak07}. Another interpretation is that it is Comptonized 
emission from a warm, optically-thick corona covering the inner part 
of the disk \citep{rob07}. A second subgroup of ULXs (including NGC\,1365 X1 
in 2004) sometimes show a thermal soft excess ($kT_{\rm in} \approx 0.15$ keV) 
in addition to the power-law component; but the soft excess
contributes only $\la 30$\% of the luminosity \citep{sto06}. 
Again, this is not evidence of a canonical high/soft state, 
which is by definition dominated by the disk. In fact, 
it is not uncommon also for Galactic BHs in the low/hard state 
to have a non-dominant soft excess from the truncated outer disk 
in addition to the hard power-law component \citep{rem06}.
In summary, although there are two subgroups of ULXs with possible 
evidence of thermal components in their X-ray spectra, neither case 
corresponds to the canonical high/soft state.

\begin{figure}
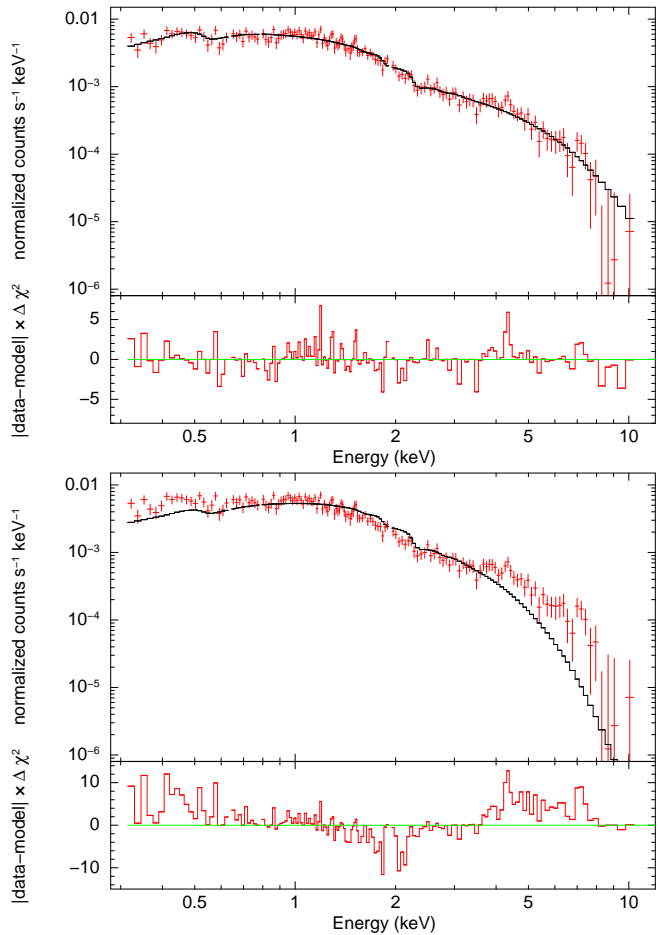

\includegraphics[angle=270,width=1\columnwidth]{fig4.ps}\\
\includegraphics[angle=270,width=1\columnwidth]{fig5b.ps}
\figcaption{Top panel: combined 2007 {\it XMM-Newton}/EPIC spectrum 
of X1, and $\chi^2$ residuals. The model is an absorbed 
power-law.  The best-fitting parameters are 
listed in Table 2. We have not included in the fit 
three ``bad'' channels (at $\approx 0.6$, $0.8$ and $2$ keV) 
that were clearly very discrepant from any fitting model, 
leading to an additional $\Delta \chi^2 \approx 30$; 
we are unable to pin down the reason for the odd behavior 
of those three channels, but we suspect they are 
related to the EPIC-pn behavior at the edge of the chip gap 
(they are not discrepant in the MOS). 
Including those three channels does not change 
the best-fitting parameters. Bottom panel: the same spectrum
cannot be well fitted by a disk-blackbody model, which 
highlights the lack of spectral curvature.}
\label{fig:figure5}
\vspace{0.3cm}
\end{figure}

\subsection{A different accretion state or a different class of BHs?}

Based on the previous arguments, we sketch three possible scenarios  
to explain the spectral state of hard power-law ULXs such as X1 and X2,
and their variability (or lack of) between states.\\
{\it a) Intermediate-mass BHs in the low/hard state}. 
In this scenario, the hard power-law spectra are attributed 
to a true low/hard state. If low/hard to high/soft state transitions 
(that is, between 
a non-thermal, radiatively-inefficient and a thermal, 
radiatively-efficient accretion mode) must always occur 
for all accreting BHs at $L_{\rm X} \sim $ a few per cent 
of $L_{\rm Edd}$, this scenario implies  
that ULXs with a hard ($\Gamma \la 2$) power-law spectrum 
and luminosities $\sim 10^{40}$ erg s$^{-1}$ 
must be powered by intermediate-mass BHs with masses $\ga$ a few $10^3 M_{\odot}$ 
\citep{win06}. In this scenario, ULXs represent a different 
physical class of BHs; the reason they never switch 
to the high/soft state is that the accretion rate from the donor 
star in never high enough. However, there is still little 
supporting evidence or theoretical justification 
for such intermediate-mass BHs. Hence, we argue that 
this is the least likely interpretation, at this stage.\\


{\it b) Stellar-mass BHs in the super-Eddington state(s)}. In this scenario, 
power-law ULXs (with or without soft excess) and convex-spectra ULXs 
are all varieties of the very high state or super-Eddington regime, 
characterized by a mix of Compton scattering, photon trapping, slim-disk 
energy advection and massive radiation-driven outflows (with 
possible transitions between substates, as seen for example 
in IC\,342 X1: \citet{kub01}).
This implies BH masses $\sim 10 M_{\odot}$, isotropic 
X-ray luminosities $\sim 1$--$30 L_{\rm Edd}$, 
and mass accretion rates $\dot{m} \sim$ a few to $\sim 1000$.
In this scenario, ULXs represent a different accretion state 
of normal stellar-mass BHs; their canonical high/soft state 
is not observed because their accretion rate is always 
above Eddington, and so is their luminosity.\\


{\it c) Massive stellar BHs without a high/soft state}. 
In this scenario 
(somewhat intermediate between the other two), BHs in ULXs 
have masses $\sim 30$--$100 M_{\odot}$ (at the upper end 
of the high-mass X-ray binary distribution), 
and isotropic luminosities 
$L_{\rm X} \sim 0.1$ to a few $L_{\rm Edd}$. 
The main observational motivation for this 
massive-stellar-BH scenario comes from 
the observed downturn in the ULX luminosity distribution 
at $L_{\rm X} \approx 2 \times 10^{40}$ erg s$^{-1}$ 
\citep{gri03,swa04}, which can be interpreted as $\sim$ 
the characteristic Eddington luminosity for the ULX population.
This implies that the isotropic luminosity 
of most ULXs is below Eddington. In turns, 
this means that ULXs must follow a different canonical sequence 
of state transitions than Galactic BHs, because 
they are always dominated by Comptonized emission 
at luminosities below, near and 
slightly above Eddington, without a ``naked-disk'' 
high/soft state in between. If there is a direct disk component, 
it is always less important than the Comptonized component, or it may 
be a slim-disk component for the most luminous sources.
Following this interpretation, there is no longer 
a gap between low/hard and very high state, just a continuous 
range of coronal optical depths (getting thicker at higher $\dot{m}$) 
and temperatures (getting cooler); one ``state'' can morph 
directly into the other.

\subsection{Two-component accretion flows?}

We argue that the scenario c) above is the most consistent 
with multiband observations, although there may also be some ULXs 
in the highly super-Eddington state of scenario b). 
One issue that can differentiate the two scenarios 
is the relative change in $\dot{m}$ necessary to produce 
a change in luminosity. For scenario c), changes in luminosity 
between $\sim 0.1$ and $\sim 1 L_{\rm Edd}$ can correspond 
to similar relative changes in the accretion rate. For scenario b), 
the radiative luminosity is strongly saturated above Eddington: 
$L \sim (1+\ln \dot{m})$; changes in luminosity between 
$\sim 1$ and $\sim 10 L_{\rm Edd}$ require an increase in 
$\dot{m}$ by several orders of magnitude, which may severely constrain 
the mechanism responsible for the transition, and may have stronger 
detectable effects on the optical counterpart, or outflow density.

Why would ULXs lack a canonical high/soft state? 
Their BHs would be a few times, or perhaps even 
an order of magnitude more massive than Galactic BHs---but this should 
not directly affect their state-transition properties. 
We suggest that the type of donor stars 
and the mechanism of mass transfer can cause 
a different spectral-state behaviour in Galactic BHs and ULXs.
The canonical spectral state classification 
is based mostly on the appearance of Galactic BHs 
in low-mass X-ray binaries, also known as soft X-ray transients.
Instead, the most luminous ULXs are thought to have 
OB donors, with masses $\ga 10 M_{\odot}$, 
based on theoretical and observational considerations. 
Stellar evolution models \citep{rap05}
show that OB stars can provide the required mass transfer rates 
$\ga 10^{-6} M_{\odot}$ yr$^{-1}$ over their nuclear timescale 
$\sim 10^6$ yr, with a peak as the donor evolves 
to the blue-supergiant stage. And optical observations 
of ULX counterparts show that they are indeed predominantly located in  
OB associations or regions of recent star formation, 
even though it is always difficult to identify 
the donor star, and separate the optical contribution 
of the donor star from that of the irradiated 
accretion disk \citep{ram06,gri08}. 
A typical ULX donor star is also likely to be filling 
its Roche lobe, because the fraction of gas captured 
by the BH from wind accretion alone may not 
be sufficient to produce the observed X-ray 
luminosity. 
We speculate that ULX donor stars may provide both 
high-angular-momentum gas, feeding the disk 
through Roche-lobe overflow, and low-angular-momentum 
gas, feeding the Comptonizing medium and jet through 
a sub-Keplerian wind, perhaps enhanced by the X-ray 
irradiation of the secondary. This would be analogous 
to the two-flow scenario proposed for some Galactic BHs 
with massive donors \citep{smi02}. 
In fact, it may not be a coincidence that Cyg X-1, 
perhaps the closest Galactic analog to a ULX in terms 
of accretion geometry and donor star, has only rarely been seen  
in a true high/soft state \citep{dot97}; 
instead, most of its state transitions are between 
the low/hard state (where it spends the majority of its time) 
and what should be properly labelled as very high state, 
with radio activity and an X-ray spectrum consisting 
of a strong power-law plus thermal component \citep{rem06,fen03,zha97,zha97b}.  
We note as another possible analogy that the accretion mode 
of radio galaxies (as indicated by their emission-line 
type and radio power) may depend on their source 
of gas (Bondi accretion of the hot gas and/or disk accretion 
of cold gas), not simply on the total accretion rate \citep{har07}.

The arguments above may at least qualitatively explain why ULXs 
can retain a hard power-law spectrum even 
at X-ray luminosities $\sim 0.1$--$1 L_{\rm Edd}$. 
The other factor that may determine the ULX state behaviour 
is their high accretion rate, as mentioned earlier. BH transients 
can switch to a soft, disk-dominated state 
during an outburst only if $\dot{m} \la 1$; in that case, the accretion flow 
can settle into a quasi-steady standard-disk configuration.
But this cannot occur if the accretion rate during their outbursts 
stays persistently above Eddington 
(the inner part of the disk cannot survive at $\dot{m} > 1$). 
We suggest that some transient ULXs may switch directly 
from the low/hard state (at low accretion rates) to the 
slim-disk state (which produces a convex X-ray spectrum), 
or to the Compton-scattering-dominated very high state 
(which produces a cut-off power-law spectrum).
The standard disk may only survive or be directly visible  
at large radii, and may produce the cool soft excess 
\citep{rob07}.
In this scenario, the reason why ULXs have long-lasting 
phases of super-Eddington mass transfer rate may be due  
to the different mass or evolutionary stage of their donor stars, 
compared with the typical donor stars of Galactic BH transients.

\begin{figure}
\includegraphics[angle=270,width=1\columnwidth]{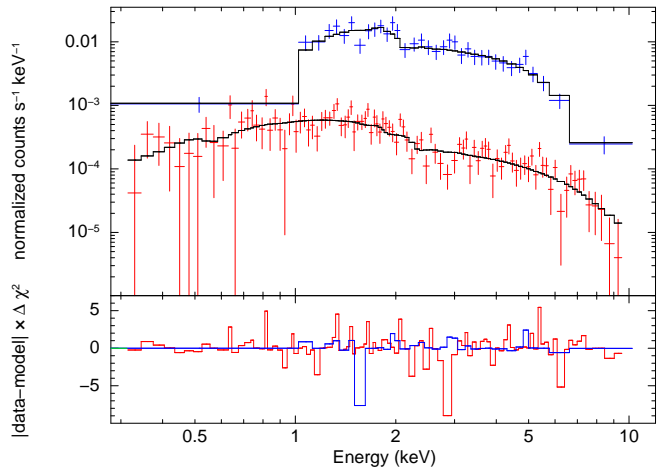}
\figcaption{Red datapoints (lower count rate): combined 
{\it XMM-Newton}/EPIC spectrum of X2 from 2007 Jun--Jul, 
with $\chi^2$ fit residuals. 
Blue datapoints (higher count rate): {\it Chandra}/ACIS spectrum 
of X2 from 2006 April 23, with $\chi^2$ fit residuals.
For both spectra, the best-fitting model is an absorbed 
power-law (Table 3).}
\label{fig:figure6}
\vspace{0.3cm}
\end{figure}

\subsection{Thresholds for the low/hard state}

To make progress on these issues, we need to understand 
more about the threshold luminosity and accretion 
rate at which BHs (including Galactic X-ray binaries) 
switch from the low/hard to the high/soft state, 
and to determine whether, in some cases, 
a system can remain in a hard state up 
to luminosities $\sim 0.1$--$1 L_{\rm Edd}$, 
and then evolve directly towards the very high state.
The scenario of a universal threshold between low/hard and high/soft 
state at $\dot{m} \sim 0.01$ works if the low/hard state 
consists of a radiatively-inefficient, advection-dominated 
flow \citep{esi97}. Such flow is not expected to survive 
at higher accretion rates, when radiative cooling 
becomes more efficient.
However, other accretion models may explain 
the persistence of a hard state up to higher luminosities: 
for example a ``luminous hot accretion flow''
\citep{yua04} may be consistent with hard-state 
luminosities up to at least $0.1 L_{\rm Edd}$.
Further investigations should determine whether 
the hard state in ULXs (and perhaps in some classes of AGN) 
can extend to even higher luminosities, 
if the dominant power-law component is produced 
by a magnetized corona or a jet, perhaps favored 
by sub-Keplerian accretion flows.
Observationally, low/hard to high/soft transitions 
during several outbursts of the Galactic BHs GX\,339-4 and 
XTE J1550$-$564 have been seen at different 
X-ray luminosities, sometimes as high 
as $\approx 0.1 L_{\rm Edd}$ \citep{yuw07}.
The existence of a linear relation between 
the peak luminosity in the low/hard and high/soft states, 
and the properties of the time delay between the two 
states, have been interpreted \citep{yuw07} 
as possible evidence of the two-flow accretion geometry 
(sub-Keplerian flow responsible for the low/hard 
emission, and Keplerian flow feeding the disk).
Hard-state luminosities $\ga 10^{38}$ erg s$^{-1}$ 
have also been seen in the Galactic BH candidate Cyg X-3 
\citep{hja08}, which has a Wolf-Rayet donor with a strong wind.
Finally, there is a group of at least 8 Galactic BHs 
\citep{bro04} that remain in the low/hard state 
throughout their outbursts. Understanding why 
those BHs never switch to the high/soft state
will also provide a useful comparison for the ULX behavior.


If our suggestion that ULXs lack a canonical high/soft state 
is correct, we should gain further insight 
by studying their radio behaviour.
Accreting BHs have a steady jet (flat-spectrum radio-core emission) 
in the low/hard state; instead, they are radio-quiet 
in the high/soft state, when the jet is quenched \citep{fen04}, 
and radio-flaring in the very high state. 
If some BHs never switch to the high/soft state, 
it will be interesting to determine whether 
they always have a jet, which may carry out a substantial 
fraction of the accretion power; perhaps the only transition 
is between a steady jet at lower luminosities and a flaring jet 
at higher luminosities. 
So far, radio observations have not yet been sensitive enough 
to detect point-like radio core emission in ULXs, and monitor 
their short-term variability in relation to their X-ray state. 
Only the much brighter, persistent, extended lobe emission 
(optically-thin synchrotron nebulae) has been identified 
in some ULXs \citep{mil05,lan07}.

\begin{acknowledgments}
We thank the referee for his/her suggestions on the significance 
of the slim-disk state, which allowed us to improve the Discussion. 
RS acknowledges support from a Leverhulme Early-Career Fellowship, 
a UK-China Fellowship for excellence, and from Tsinghua University 
(Beijing); he also thanks Prof.~Shuang-Nan Zhang and Prof.~Wenfei Yu 
for their insightful suggestions and their hospitality at Tsinghua 
University and at the Shanghai Observatory, respectively. 
GR acknowledges support from his NASA-G06-7102X and NNX07AR90G grants.
\end{acknowledgments}


\end{document}